\documentclass[a4paper,12pt]{article}
\usepackage{graphicx,amssymb,bm,latexsym}
\newcommand{\gsim}{\, \mbox{\raisebox{-1.ex}
{$\stackrel{\textstyle>}{\textstyle\sim}$}}\,}
\newcommand{\lsim}{\, \mbox{\raisebox{-1.ex}
{$\stackrel{\textstyle<}{\textstyle\sim}$}}\,}

\newcommand{\bea}{\begin{eqnarray}}
\newcommand{\ena}{\end{eqnarray}}
\newcommand{\vs}[1]{\vspace{#1 mm}}

\renewcommand{\a}{\alpha}
\renewcommand{\b}{\beta}
\renewcommand{\c}{\gamma}

\newcommand{\e}{\epsilon}

\newcommand{\dsl}{\pa \kern-0.5em /}

\newcommand{\pa}{\partial}

\newcommand{\nn}{\nonumber\\}
\newcommand{\p}[1]{(\ref{#1})}

\begin{document}

\renewcommand{\thefootnote}{\fnsymbol{footnote}}
\begin{titlepage}

\begin{flushright}
WU-AP/185/04\\
OU-HET 473 \\
hep-th/0405205
\end{flushright}

\vs{10}
\begin{center}
{\large\bf Inflation from
M-Theory with Fourth-order Corrections and Large Extra Dimensions}
\vs{15}

{\large
Kei-ichi Maeda$^{a,b,c}$\footnote{e-mail address:
maeda@gravity.phys.waseda.ac.jp} and Nobuyoshi Ohta$^d$\footnote{e-mail
address: ohta@phys.sci.osaka-u.ac.jp}}\\
\vs{10}
$^a${\em Department of Physics, Waseda University,
Shinjuku, Tokyo 169-8555, Japan}\\
$^b$ {\em Advanced Research Institute for Science and Engineering,
Waseda University, Shinjuku, Tokyo 169-8555, Japan~} \\
$^c$ {\em Waseda Institute for Astrophysics, Waseda University,
Shinjuku, Tokyo 169-8555, Japan~}  \\
$^d${\em Department of Physics, Osaka University,
Toyonaka, Osaka 560-0043, Japan}

\end{center}
\vs{15}
\centerline{{\bf{Abstract}}}
\vs{5}

We study inflationary solutions in the M-theory.
Including the fourth-order curvature correction terms, we find three
generalized de Sitter solutions, in which our 3-space expands exponentially.
Taking one of the solutions, we propose an inflationary scenario of the
early universe. This provides us a natural explanation for
large extra dimensions in a brane world, and suggests some connection
between the 60 e-folding expansion of inflation and TeV gravity based on
the large extra dimensions.

\end{titlepage}
\newpage
\renewcommand{\thefootnote}{\arabic{footnote}}
\setcounter{footnote}{0}
\setcounter{page}{2}

\section{Introduction}

Recently there has been renewed interest in obtaining accelerating
universe from higher-dimensional gravitational theories.
The study of this subject gets its urgency from the recent discovery of
the accelerated expansion of the present universe as well as the
confirmation of the existence of the early inflationary epoch~\cite{WMAP}.
Though it is not difficult to construct cosmological models with these
features, it is desirable to derive such a model from the fundamental
theories of particle physics that incorporate gravity. The most promising
candidate for such theories are the ten-dimensional superstrings or
eleven-dimensional M-theory, which are hoped to give models of accelerated
expansion of the universe upon compactification to four dimensions.

It has been shown that models with certain period of accelerated
expansion can be obtained from the higher-dimensional vacuum Einstein
equation if one takes the internal space hyperbolic and its size depending
on time~\cite{TW}. It has been shown~\cite{NO} that this class of  models
is obtained from what are known as S-branes~\cite{Sbrane1,Sbrane2}
in the limit of vanishing flux of three-form fields (see also~\cite{Sbrane3}).
It is found that this wider class of solutions gives accelerating
universes for internal spaces with zero and positive curvatures as well
if the flux is introduced. Further discussion
of this class of models has been given in refs.~\cite{W}-\cite{cosm3}.

It turns out, however, that the model thus obtained does not give enough
e-folds necessary to explain the cosmological problems such as
horizon and flatness problems~\cite{NO,W}.
Some efforts to overcome this problem were made in the present framework
in ref.~\cite{cosm3}.

We note that the scale when the acceleration occurs in this type
of models is basically governed by the Planck scale in the higher (ten
or eleven) dimensions. With phenomena at such high energy, it is expected
that we cannot ignore higher order corrections in the theories at least
in the early universe. In fact there are terms of higher orders in the
curvature to the lowest effective supergravity action coming from
superstrings or M-theory. In four dimensions,
many studies have been done with such correction terms~\cite{old1}.

The cosmological models in higher dimensions were also studied intensively
in the 80's by many authors~\cite{KK_cosmology,ISHI}.
Among them, the model with Gauss-Bonnet term is very interesting for
the above reason~\cite{ISHI}.
It was shown that there are two exponentially expanding solutions,
which may be called generalized de Sitter solutions since the size of the
internal space depends on time (otherwise there is no solution of this type).
In both solutions, the external space inflates, while the internal space
shrinks exponentially. (There are also two time-reversed solutions, i.e. the
external space shrinks exponentially but the internal space inflates.)
One solution is stable and the other is unstable. Since the present
universe is not in the phase of de Sitter expansion with this energy scale,
we cannot use the stable solution for a realistic universe. If we adopt
the unstable solution, on the other hand, we may not find sufficient
inflation unless we fine-tune the initial values. The higher-order
curvature terms called Lovelock gravity were also considered in
higher-dimensional cosmology~\cite{DF}.

However, most of the work considered powers of scalar curvature or Lovelock
gravity, which are not the types of corrections arising in type II
superstrings or M-theory. In particular, it is known that the coefficient of
the Gauss-Bonnet terms vanishes and the first higher order corrections start
with $R^4$ terms (one is the fourth order Lovelock gravity and the
other contains higher derivatives)~\cite{TBB}.
The purpose of this paper is to examine how these corrections in the
fundamental theories modify the above cosmological models and
whether we can get interesting cosmological scenario with large e-folds.
We focus on the solutions to the vacuum Einstein equations with these higher
order corrections in this paper since the basic feature can be obtained
in this simple setting.

We find interesting models with power law as well as exponential expansions.
These seem to give enough e-folds, and the solutions suggest that the
fundamental Planck scale is rather small as $\lsim 10^3$ TeV and the size
of internal space grows rather large with the scale $\lsim 10$ TeV$^{-1}$.
These may provide the models of large extra dimensions discussed by
Arkani-Hamed et al.~\cite{arkani}. We have found similar results for
several superstrings and M-theory, but here we present only the results of
exponential expansions for M-theory, leaving other details to
a separate paper~\cite{MOII}.

\section{Vacuum Einstein Equations with $R^4$ corrections}

We consider the low-energy effective action for the M-theory:
\bea
S &=& S_{\rm EH} + S_4,
\ena
where
\bea
S_{\rm EH} &=& \frac{1}{2\kappa_{11}^2} \int d^{11} x \sqrt{-g} R,\\
S_4 \! &=&  \frac{1}{2\kappa_{11}^2}\int d^{11} x
\sqrt{-g}\left[\c  \tilde{J}_0
+2 \b \tilde{E}_8\;\!\right] \,.
\label{4th}
\ena
\bea
\tilde{E}_8&=&-{1\over 2^4 \times 3!}
\e^{\a\b\c\mu_1 \nu_1 \ldots \mu_4 \nu_4}
\e_{\a\b\c\rho_1 \sigma_1 \ldots \rho_4 \sigma_4} R^{\rho_1
\sigma_1}{}_{\mu_1 \nu_1}
\cdots R^{\rho_4 \sigma_4}{}_{\mu_4 \nu_4}\,,
\\
\tilde{J}_0&=&
R^{\lambda\mu\nu\kappa}R_{\a\mu\nu\b}R_{\lambda}{}^{\rho\sigma\a}
R^\b{}_{\rho\sigma\kappa}
+\frac12 R^{\lambda\kappa\mu\nu}R_{\a\b\mu\nu}R_{\lambda}{}^{\rho\sigma\a}
R^\b{}_{\rho\sigma\kappa}.
\ena
Here we have dropped contributions from forms, $\kappa_{11}^2$ is
an eleven-dimensional (11D) gravitational constant, and we leave
the coefficients $\b$ and $\c$ free although we know them for
the M-theory~\cite{TBB} as
\bea
\beta = - {\kappa_{11}^2 ~T_2\over 3^2\times 2^{10} \times (2\pi)^4} ,\qquad
\gamma = - {\kappa_{11}^2 ~ T_2\over 3 \times 2^{4}\times (2\pi)^4} \,,
\ena
where $T_2=({2\pi^2 /\kappa_{11}^2})^{1/3}$ is the membrane tension.
Type II superstring has the same couplings in 10 dimensions,
so we can discuss this case if we keep the dilaton field constant,
but we consider 11D theory in this paper.
Here we should note that contributions of the Ricci tensor $R_{\mu\nu}$
and scalar curvature $R$ are not included in the fourth-order
corrections~(\ref{4th}) because these terms are not uniquely fixed.

Since we are interested in a cosmological time-dependent solution,
we take the metric of our spacetime as
\bea
ds_{11}^2 &=& -e^{2u_0(t)} dt^2 + e^{2u_1(t)} \sum_{i=1}^3 (dx^i)^2
+ e^{2u_2(t)} \sum_{a=5}^{11} (dy^a)^2\,,
\label{met1}
\ena
where we assume that the external 3-space and the internal 7-space
are flat. Taking variation of the action with respect to $u_0$,
$u_1$, and $u_2$, we obtain three basic equations, whose explicit forms
will be given in a forthcoming paper~\cite{MOII}.

\subsection{Generalized de Sitter Solutions}

In cosmology, de Sitter inflationary expanding spacetime
is the most important solution in the early universe.
Hence, let us first look for such solutions.
Assuming the metric form of a generalized de Sitter spacetime as
\bea
u_0=0 ,\,\,u_1=\mu t ,\,\,u_2=\nu t ,\,\,
\ena
where $\mu$ and $\nu$ are some constants,
we obtain three algebraic equations:
\bea
\label{1}
&& \mu^2 + 7\mu\nu + 7\nu^2
 + 20160 \b  \mu \nu^5
\Big[ 7 \mu^2 + 7 \mu \nu +  \nu^2 \Big]  \nn
&& -7 \c \Big[ 12 \mu^8 + 7\mu^2 \nu^2 (\mu^2+\nu^2+\mu\nu )^2
+168 \nu^8 +7 \mu^4 \nu^2(2\mu+\nu)^2  +21\mu^2 \nu^4(\mu+2\nu)^2  \Big]
\nn && + 4\c (3\mu+7\nu)\Big[ 6 \mu^7 + 42 \nu^7 + 7\mu^2 \nu^2 (\mu+\nu)
(\mu^2+\nu^2+\mu\nu ) \Big]
=0,  \\ \nn
\label{2}
&& 3 \mu^2 + 14\mu\nu + 28 \nu^2  + 20160 \b \nu^5\Big[ 6 \mu^3 + 24
\mu^2 \nu + 14 \mu\nu^2+  \nu^3 \Big]
\nn
&& +3 \c \Big[ 12 \mu^8 + 7\mu^2 \nu^2
(\mu^2+\nu^2+\mu\nu )^2
 +  168 \nu^8 + 7\mu^4 \nu^2(2\mu+\nu)^2  + 21\mu^2 \nu^4(\mu+2\nu)^2
\Big]
\nn
&& -2 \c \mu (3\mu+7\nu)\Big[ 48 \mu^6 + 7 \nu^2 (3\mu^2  +2\mu\nu +\nu^2)
(\mu^2+\nu^2+\mu\nu ) \nn
&&~~~~~~~~~~~~~~~~~~~~ +14\mu^2 \nu^2(2\mu+\nu)(3\mu+\nu)  +
42 \nu^4(\mu+2\nu)(\mu+\nu) \Big] \nn
&& +2\c \mu (3\mu+7\nu)^2\Big[ 12 \mu^5+7 \nu^2(\mu+\nu)  (\mu^2+\nu^2
+\mu\nu )
\Big]
 =0,
\\ \nn
\label{3}
&& 2\mu^2 + 6\mu\nu + 7 \nu^2 + 2880\b \mu\nu^4 \Big[
15\mu^3 + 46 \mu^2 \nu   +38\mu\nu^2 + 6 \nu^3
\Big] \nn
&& + \c \Big[ 12 \mu^8 + 7 \mu^2 \nu^2(\mu^2+\nu^2+\mu\nu )^2  +168
\nu^8 +7\mu^4 \nu^2(2\mu+\nu)^2    +21\mu^2 \nu^4(\mu+2\nu)^2  \Big]
\nn
&& -2 \c \nu (3\mu+7\nu)\Big[ 96 \nu^6  +
\mu^2   (\mu^2+2\mu\nu+3\nu^2)(\mu^2+\nu^2+\mu\nu )
\nn &&
~~~~~~~~~~~~~~~~~~~~+2\mu^4  (2\mu+\nu)(\mu+\nu) +6\mu^2
\nu^2(\mu+2\nu)(\mu+3\nu)
\Big]
\nn && +2\c\nu (3\mu+7\nu)^2\Big[ 12 \nu^5+\mu^2  (\mu+\nu) (\mu^2+\nu^2
+\mu\nu )
\Big]=0 \,.
\ena

\begin{figure}[tb]
\centering
\includegraphics[width=16cm]{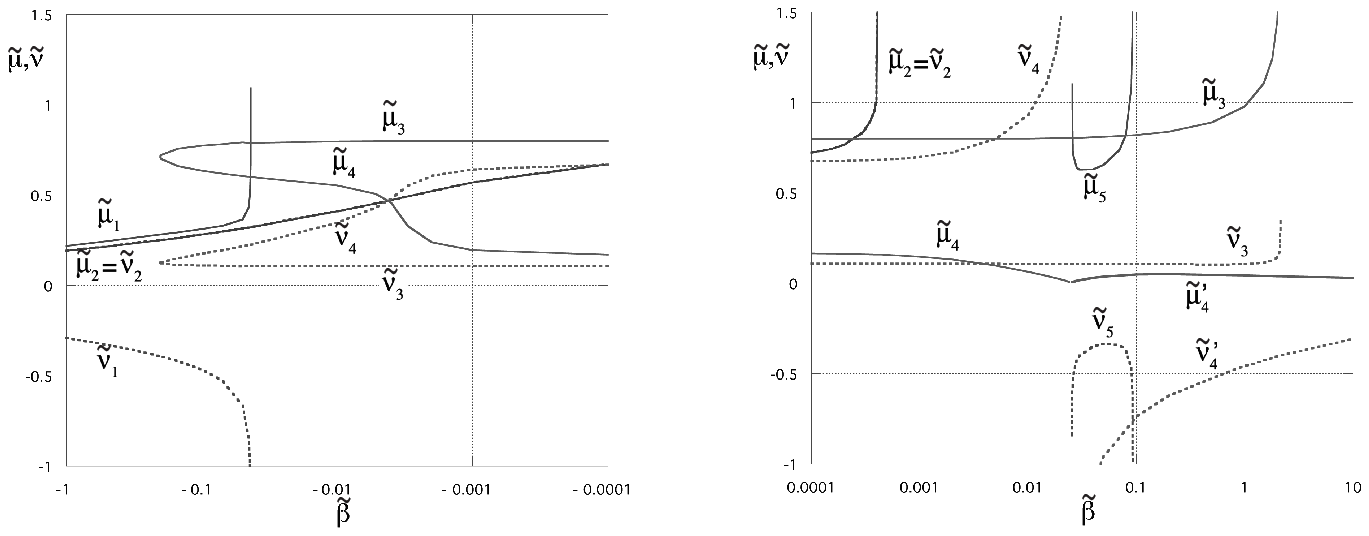}
\caption[fig1]{Five generalized de Sitter solutions ${\rm N}_i
(\tilde{\mu}_i,\tilde{\nu}_i)$ ($i=1\sim 5$) with
$\tilde{\mu}_i \geq 0$ with respect to $\tilde{\b}$ for $\c<0$.
Each pair of points $(\tilde{\mu}_i,\tilde{\nu}_i)$ with the same
value of $\tilde{\b}$ gives one solution. Number of solutions changes
with the value of $\tilde{\b}$.
There is another set of time-reversed solutions ${\rm N}'_i$ with
$\tilde{\mu}_i \leq 0$.
}
\label{fig1}
\end{figure}

\begin{figure}[tb]
\centering
\includegraphics[width=10cm]{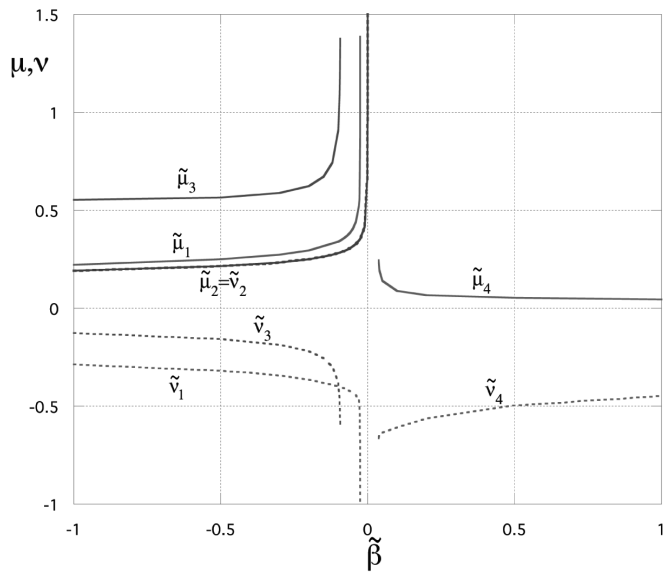}
\caption[fig2]{Four generalized de Sitter solutions ${\rm P}_i
(\tilde{\mu}_i,\tilde{\nu}_i)$ ($i=1\sim 4$) with
$\tilde{\mu}_i > 0$ with respect to $\tilde{\b}$ for $\gamma>0$.
Each pair of points $(\tilde{\mu}_i,\tilde{\nu}_i)$ with
the same value of $\tilde{\b}$ gives one solution.
Number of solutions changes with the value of $\tilde{\b}$.
There is another set of time-reversed solutions ${\rm P}'_i$ with
$\tilde{\mu}'_i < 0$.
No solution exists for $\gamma>0$ and $\beta=0$.}
\label{fig2}
\end{figure}

Since these equations are very complicated, we have solved them numerically.
If $\c$ does not vanish, rescaling $\b$, $\c$, $\mu$ and $\nu$ as
\bea
\tilde{\b}=\b /|\c| \, \,,
\tilde{\c}=\c /|\c| ~(=1 ~{\rm or}~ -1)  \, \,,
\tilde{\mu}= \mu |\c|^{1/6}  \,, \,{\rm and}\,\,
\tilde{\nu}= \nu |\c|^{1/6}\,,
\ena
we can always set $\c$ to $-1$ if it is negative (or 1 if positive).
We also have to rescale time coordinate as $\tilde{t}=|\c|^{-1/6} t$.
The typical dynamical time scale is then given by $|\c|^{1/6}\sim
0.181818 m_{11}^{-1}$, where $m_{11}=\kappa_{11}^{-2/9}$ is
the fundamental Planck scale. After this scaling, we have only one
free parameter $\tilde{\b}$.
In Fig.~\ref{fig1}, we depict numerical solutions
N$_i(\tilde{\mu}_i,\tilde{\nu}_i)$ ($i=1\sim 5$) with $\tilde{\mu}_i \geq 0$
with respect to $\tilde{\b}$ for the case of $\gamma<0$.
We note that there are always time-reversed solutions
N$'_{i}(\tilde{\mu}'_i,\tilde{\nu}'_i)$ ($i=1\sim 5$) obtained by
$(\tilde{\mu}'_i,\tilde{\nu}'_i)=-(\tilde{\mu}_i,\tilde{\nu}_i)$ which
are not shown explicitly. We find that M-theory
($\tilde{\c}=-1,\tilde{\b}=\tilde{\b}_S=-1/(3\times 2^6) \approx -0.0052083$)
has three solutions
\bea
\label{oursol}
{\rm N}_2 (\tilde{\mu}_2,\tilde{\nu}_2) &=& (0.45413,0.45413),\nn
{\rm N}_3 (\tilde{\mu}_3,\tilde{\nu}_3) &=& (0.79802,0.10781),\\
{\rm N}_4 (\tilde{\mu}_4,\tilde{\nu}_4) &=& (0.50754,0.43025).\nonumber
\ena
If $\gamma>0$, we find three solutions P$_i (i=1\sim 3)$
for  $\beta<0$, while just one P$_4$ for
$\beta>0$. We depict these solutions in Fig.~\ref{fig2} (for $\tilde \c=1$).
There is no solution for $\beta=0$.
In Table \ref{table_1}, we summarize their properties.
\begin{table}[bt]
\caption{Generalized de Sitter solutions ${\rm N}_i
(\tilde{\mu}_i,\tilde{\nu}_i)$  ($i=1\sim 5$) and  ${\rm
P}_i (\tilde{\mu}_i,\tilde{\nu}_i)$  ($i=1\sim 4$)  with
$\tilde{\mu}_i \geq 0$ for various values of $\tilde{\b}$ in the cases of
$\tilde{\gamma}=-1$ and of $\tilde{\gamma}=1$, respectively.
The solution ${\rm N}'_4$ is the partner of the solution
${\rm N}_4$ with $\tilde{\mu}_4<0$, which is found when it is extended
to the range of $\tilde{\beta}>0.025668$.
Five eigen modes for linear perturbations are also shown.
($m$s,$n$u) means that there are $m$ stable modes and
$n$ unstable modes. The solution has many stable modes
if its 10-volume expansion rate ($3\tilde{\mu}+7\tilde{\nu}$)
is positive.} ~\\
\begin{tabular}{|c||c||c|c|c|c|}
\hline
$\c$&solution&property&range&stability&$3\tilde{\mu}+7\tilde{\nu}$\\
\hline
\hline
&${\rm
N}_1$&$\tilde{\mu}_1>0>\tilde{\nu}_1$&$\tilde{\beta}<-0.04342$&(1s,4u)&$-$\\
\cline{2-6}
&\raisebox{-1.5ex}[0pt]{${\rm
N}_2$}&\raisebox{-1.5ex}[0pt]{$\tilde{\mu}_2=
\tilde{\nu}_2>0$}&
$\tilde{\beta}<-0.00418$&(4s,1u)&$+$\\
\cline{4-6}
&&&
$-0.00418<\tilde{\beta}<0.0004464$&(5s,0u)&$+$\\
\cline{2-6}
$\tilde{\c}=-1$&${\rm N}_3$&$\tilde{\mu}_3>\tilde{\nu}_3>0$&
$-0.202867<\tilde{\beta}<2.141$&(4s,1u)&$+$\\
\cline{2-6}
&\raisebox{-1.5ex}[0pt]{${\rm N}_4$}&$\tilde{\mu}_4>\tilde{\nu}_4>0$&
$-0.202867<\tilde{\beta}<-0.00418$&(5s,0u)&$+$\\
\cline{3-6}
&&$\tilde{\nu}_4>\tilde{\mu}_4>0$&
$-0.00418<\tilde{\beta}<0.025668$&(4s,1u)&$+$\\
\cline{2-6}
&${\rm N}'_4$&$\tilde{\mu}_4'
>0>\tilde{\nu}_4'$&
$\tilde{\beta}>0.025668$&(1s,4u)&$-$\\
\cline{2-6}
&${\rm N}_5$&$\tilde{\mu}_5>0>\tilde{\nu}_5$&
$0.025668<\tilde{\beta}<0.09247$&(1s,4u)&$-$\\
\hline
\hline
&${\rm P}_1$&$\tilde{\mu}_1>0>\tilde{\nu}_1$&&(1s,4u)&$-$\\
\cline{2-3}
\cline{5-6}
\raisebox{-1.5ex}[0pt]{$\tilde{\c}=1$}&${\rm
P}_2$&$\tilde{\mu}_2=\tilde{\nu}_2>0$&
$\tilde{\beta}<0$&(4s,1u)&$+$\\
\cline{2-3}
\cline{5-6}
&${\rm P}_3$&$\tilde{\mu}_1>0>\tilde{\nu}_1$&
&(4s,1u)&$+$\\
\cline{2-6}
&${\rm P}_4$&$\tilde{\mu}_1>0>\tilde{\nu}_1$&
$\tilde{\beta}>0$&(1s,4u)&$-$\\
\hline
\hline
\end{tabular}
\label{table_1}
\end{table}

If $\gamma=0$ and $\beta\neq 0$, we can always set $\b$ to $-1$ if it is
negative (or 1 if positive), by rescaling $\b$, $\mu$ and $\nu$ as
$
\bar{\b}=\b /|\b| ~(=1 ~~{\rm or}~~-1)\, \,,
\bar{\mu}= \mu |\b|^{1/6}  \,, \,{\rm and}\,\,
\bar{\nu}= \nu |\b|^{1/6}\,.
$
We then find two solutions
[$(\bar{\mu},\bar{\nu})=(0.22046,-0.28771), (0.19168,0.19168)$] for
$\bar{\beta}=-1$, while just one
[$(\bar{\mu},\bar{\nu})=(0.04467,-0.45111)$] for
$\bar{\beta}=1$.
There is no solution for $\b=\c=0$.

\subsection{Stability}

Since the solutions obtained above correspond to fixed points in our dynamical
system, we have to analyze their stabilities in order to see which
solutions are more generic. We have performed a linear perturbation
analysis around those fixed points.  Setting
$d{u}_1/d\tilde{t}=\tilde{\mu}_i +A_i
e^{\sigma \tilde{t}}$ and
$d{u}_2/d\tilde{t}=\tilde{\nu}_i +B_i
e^{\sigma \tilde{t}}$, where
$
|A_i|,|B_i|\ll 1$, we write down the perturbation equations.
There are five modes ($\sigma=\sigma_a^{(i)},~a=1,2,\cdots, 5$) because
the basic equations for
$\dot u_1$ and $\dot u_2$ are two third-order differential equations plus
one constraint which is second order. We show the results in
Tables \ref{table_1} and \ref{table_2}. In Table \ref{table_1}, we just
give the number of stable and unstable modes. For example, there are
one stable and four unstable modes for the solution N$_1
(\tilde{\mu}_1,\tilde{\nu}_1)$. Hence this solution may not be generic
because we have many unstable modes. The M-theory
has three solutions~\p{oursol}.
Two solutions (${\rm N}_2$ and ${\rm N}_3$) have four stable and
one unstable modes. The third solution (${\rm N}_4$) has five stable
modes, which means that this solution is stable against linear
perturbations (see Table~\ref{table_2}).
\begin{table}[htb]
\caption{The eigenvalues of linear perturbation equations for $\tilde{\c}=-1$.
We show two cases of $\tilde{\b}$: $\tilde{\b}=\tilde{\b}_S$ and one for
which we find an interesting scenario.
} ~\\
{\small
\begin{tabular}{|c||c|c|c||l|}
\hline
$\tilde{\beta}$&solution&$\tilde{\mu}_i$&$\tilde{\nu}_i$&
~~~~~five eigenvalues ($\sigma_a^{(i)},\, a=1,2, \cdots 5$)\\
\hline
\hline
&${\rm N}_2$&0.45413&0.45413&$(\,-4.5893,-4.5413,-3.6670,-0.87426,
\,0.048032\,)$\\
\cline{2-5}
$ -0.0052083$&${\rm
N}_3$&0.79802&0.10781&$(\,-4.1012,-3.1488,-2.3823,-0.76653,
\,0.95268\,)$\\
\cline{2-5}
&${\rm
N}_4$&0.50754&0.43025&$(\,-4.5344,-4.4768,-3.6344,-0.89998,-0.057629\,)$\\
\hline
\hline
&${\rm N}_1$&0.28195&$-0.39104$&$(\,-10.2506,\,0.9457\pm 1.8339
i,\,1.8914,\,12.1421\,)$\\
\cline{2-5}
\raisebox{-1.5ex}[0pt]
{$-0.2025$}&${\rm N}_2$&0.25005&0.25005&$(\,-5.0572,-2.5005,-1.2502\pm
5.7723 i,\,2.5567\,)$\\
\cline{2-5}
&${\rm N}_3$&0.71567&0.12395&$(\,-3.0268,-3.0147,-1.5073\pm 3.0912
i,\,0.01217\,)$\\
\cline{2-5}
&${\rm N}_4$&0.70803&0.12661&$(\, -3.01038,-2.9986,-1.5052\pm 3.2449
i,-0.011765\,)$\\
\hline
\hline
\end{tabular}
}
\label{table_2}
\end{table}

{}From this stability analysis, one may conclude that the solution ${\rm
N}_4$ is  most preferable spacetime in this model. We have inflationary
expansion not only in 3D external space but also in 7D internal space.
The expansion rates in both spaces ($\mu, \nu$)  are, however,  almost
the same. Hence, we would find our present world in which scales of
two spaces are not so different. This solution also predicts that
inflation never ends because it is stable. For a realistic
cosmological model, the solution must be unstable because inflation should
end. On the other hand, we also want such a solution to be rather generic
which requires some sort of stability. This would be achieved if the
solution contains only one unstable mode, and then the generic spacetime may
first approach this solution and gradually leave it, recovering the
present Friedmann universe, where we expect the higher order terms become
irrelevant. We find that the solution ${\rm N}_3$
may give one possible candidate for such a model. We now discuss a new
scenario obtained from this solution in the next section.

\section{A Scenario for Large Extra Dimensions}

Let us discuss the evolution of the early universe for the solution
${\rm N}_3 (\tilde{\mu}_3,\tilde{\nu}_3)=(0.79802,0.10781)$ with
$\b=\b_{S}$. The scale factor of the external space expands as
$e^{0.79802\tilde{t}}$. For a successful inflation (resolution of flatness
and horizon problems), we need at least 60 e-foldings. Let us assume that
inflation will end after 60 e-foldings, i.e.
$0.79802  \tilde{t}_{\rm end}\approx 60$. The inflation will end  because
${\rm N}_3$ has one unstable mode. During inflation  the internal space
also expands exponentially. When inflation ends, its scale  becomes
$e^{0.10781 \tilde{t}_{\rm end}}\approx 4000$ times larger than the
initial  scale length, which we assume to be the 11D Planck length
($m_{11}^{-1}$). After inflation, if the internal space settles down
to static one, the present radius of extra dimensions is $R_0 \sim 4000
m_{11}^{-1}$. Since this is slightly larger than the fundamental scale
length, we may adopt the model of large extra dimensions, which was first
proposed as a brane world  by Arkani-Hamed et al.~\cite{arkani}. In this
model, the 4D Planck mass is given by
\bea
m_4^2\sim R_0^7 m_{11}^9 \sim  1.6\times 10^{25} m_{11}^2\,.
\label{m_4}
\ena
We then find
\bea
m_{11}\sim 2.5 \times 10^{-13} m_{4} \sim 600 {\rm TeV} \,.
\ena
This is our fundamental energy scale.
The present scale of extra dimensions is $4000 m_{11}^{-1}\sim 7$
TeV$^{-1}$, which could be observed in the accelerators of next generation.

We can also put our argument in a different way.
Suppose that the e-folding of inflation is $N$, which is related to the
stability of the solution. The 3-space expands as
$e^N=e^{\tilde{\mu}\tilde{t}_{\rm end}}$, while the internal space
becomes $e^{\tilde{\nu}\tilde{t}_{\rm end}}$ times larger. It follows from
Eq.~(\ref{m_4}) that
\bea
m_{11}\sim e^{-{7\nu \over 2\mu}N} m_4\,.
\ena
Since $m_{11}\gsim 1$ TeV from the present experiments, we have a
constraint on the e-folding as $N\lsim 10\mu/\nu$. Then if we have TeV
gravity and $\mu \gsim 6\nu (> 0)$, we can naturally explain why the
e-folding of inflation is about 60 and but not so large.
Recall that the solution ${\rm N}_3$ with $\gamma<0$ gives
$5.72<\mu/\nu<10.22$ (corresponding to $57\lsim N\lsim 102$)
for any value of $\beta$.

Although the above solution ${\rm N}_3$ has one unstable mode, its
eigenvalue $\sigma_5^{(3)}$ is of the same order of magnitude as other
eigenvalues of stable modes as seen from Table~\ref{table_2} and is
a little too large to give enough expansion.
If the eigenvalue of the unstable mode is much smaller than those of other
four stable modes, a preferable generalized de Sitter solution is naturally
obtained for a wide range of initial conditions. Can we find such a
possibility in superstrings or M-theory? We note that our starting
Lagrangian has some ambiguity, that is, the fourth-order correction
term $S_4$ is fixed up to the Ricci curvature tensors. If we include
correction terms including the Ricci curvature tensors, our basic
equations will be modified. We might effectively take their effect into
account by changing the value of our coefficient $\c$ or $\tilde \b$ after
the rescaling. Thus we may look for a preferable solution by changing
$\tilde{\b}$. We find that the solution ${\rm N}_3$ with $\tilde{\b}=-0.2025$
shows interesting behaviors (see Table~\ref{table_2}). Four modes are stable
and the eigenvalue of one unstable mode is very small, i.e.
$\sigma^{(3)}_5=0.01217$. Then the time scale in which this  unstable
mode becomes important is evaluated as
$\tilde{t}_{\rm us}\approx (\sigma^{(3)}_5)^{-1}
\sim 82$. Since the eigenvalues of other stable modes are of order unity,
for a wide range of initial conditions, general solutions first approach
the solution ${\rm N}_3$, which gives us an inflationary
stage, after one dynamical time ($\tilde{t}\sim O(1)$).
The unstable mode becomes important at $\tilde{t}\sim \tilde{t}_{\rm us}$,
and then the inflation ends. We may have
enough e-folding time of inflation ($N\approx \tilde{\mu}_3\tilde{t}_{\rm
us} \sim 58.8$). In this case, we find $m_{11}\sim 3.3\times 10^{-16} m_4
\sim 0.8$ TeV, which gives us a TeV gravity theory.

Although we find a successful exponential expansion and its natural
end, this is not enough for a successful inflation.
We need a reheating mechanism and have to create a density fluctuation
as a seed of cosmic structure.
A gravitational particle creation may provide a reheating
mechanism~\cite{reheating}, because the background spacetime is time
dependent and there might have some oscillation when the internal space
settles down to static one which is required to explain our present
universe. As for a density perturbation, our model may not give a
good scenario because our energy scale is $|\gamma|^{-1/6} \sim 5 m_{11}
\sim 4$ TeV. We have to invoke other mechanism for density perturbations
such as a curvaton model~\cite{curvaton}.

\section*{Acknowledgments}

We would like to thank Y. Hyakutake, T. Shiromizu, T. Torii, D. Wands,
M. Yamaguchi and J. Yokoyama for useful  discussions. The work was
partially supported by the Grant-in-Aid for Scientific Research Fund of
the MEXT (Nos. 14540281, 16540250 and 02041) and by the Waseda University
Grant for Special Research Projects and  for The 21st Century
COE Program (Holistic Research and Education Center for Physics
Self-organization Systems) at Waseda University.

\newcommand{\NP}[1]{Nucl.\ Phys.\ B\ {\bf #1}}
\newcommand{\PL}[1]{Phys.\ Lett.\ B\ {\bf #1}}
\newcommand{\CQG}[1]{Class.\ Quant.\ Grav.\ {\bf #1}}
\newcommand{\CMP}[1]{Comm.\ Math.\ Phys.\ {\bf #1}}
\newcommand{\IJMP}[1]{Int.\ Jour.\ Mod.\ Phys.\ {\bf #1}}
\newcommand{\JHEP}[1]{JHEP\ {\bf #1}}
\newcommand{\PR}[1]{Phys.\ Rev.\ D\ {\bf #1}}
\newcommand{\PRL}[1]{Phys.\ Rev.\ Lett.\ {\bf #1}}
\newcommand{\PRE}[1]{Phys.\ Rep.\ {\bf #1}}
\newcommand{\PTP}[1]{Prog.\ Theor.\ Phys.\ {\bf #1}}
\newcommand{\PTPS}[1]{Prog.\ Theor.\ Phys.\ Suppl.\ {\bf #1}}
\newcommand{\MPL}[1]{Mod.\ Phys.\ Lett.\ {\bf #1}}
\newcommand{\JP}[1]{Jour.\ Phys.\ {\bf #1}}

\end{document}